\documentclass[aps,prd,amsfonts,amssymb,preprint,eqsecnum,nofootinbib,superscriptaddress]
{revtex4} 
\usepackage{graphics,epsfig}

\newcommand{\pslash}{p\llap{/\kern-0.3pt}}
\newcommand{\qslash}{q\llap{/\kern-0.3pt}}
\newcommand{\rslash}{r\llap{/\kern-0.3pt}}
\newcommand{\lslash}{\ell\llap{/\kern-0.3pt}}
\newcommand{\Dbar}{$\overline{{\rm D}8}\ $}
\newcommand{\dbar}{\overline{{\rm D}8}}
\newcommand{\Tr}{{\rm Tr}}
\newcommand{\Ttilde}{\widetilde{T}}
\begin{document}
\preprint{WM-07-102}
%
% Title of paper
\title{\vspace*{0.5in}
Holographic Electroweak Symmetry Breaking from D-branes
\vskip 0.1in}
\author{Christopher D. Carone}\email[]{cdcaro@wm.edu}
\author{Joshua Erlich}\email[]{jxerli@wm.edu}
\author{Marc Sher}\email[]{mtsher@wm.edu}
\affiliation{Particle Theory Group, Department of Physics,
College of William and Mary, Williamsburg, VA 23187-8795}
\date{April 2007}
\begin{abstract}
We observe several interesting phenomena in a technicolor-like model of electroweak
symmetry breaking based on the D4-D8-\Dbar system of Sakai and Sugimoto.  The benefit
of holographic models based on D-brane configurations is that both sides of the
holographic duality are well understood.  We find that the lightest technicolor
resonances contribute negatively to the Peskin-Takeuchi $S$-parameter, but heavy
resonances do not decouple and lead generically to large, positive values of $S$,
consistent with standard estimates in QCD-like theories.  We study how the $S$ parameter and
the masses and decay constants of the vector and axial-vector techni-resonances vary
over a one-parameter family of D8-brane configurations.  We discuss possibilities
for the consistent truncation of the theory to the first few resonances and suggest
some generic predictions of stringy holographic technicolor models.
\end{abstract}
\maketitle

\section{Introduction} \label{sec:intro}
In recent years, the holographic relationship between strongly-coupled
field theories and higher-dimensional theories that include gravity (the
AdS/CFT correspondence~\cite{AdSCFT}) has allowed quantitative predictions
for observables in QCD that are in surprisingly good agreement with the
experimental data~\cite{AdSQCD1,AdSQCD2,AdSQCD3,Erdmenger,Evans}.  Encouraged by the
success of such models, a similar approach to studying technicolor-like
models of electroweak symmetry breaking (EWSB) based on the AdS/CFT correspondence
has led to new possibilities for physics that may be probed at the Large Hadron
Collider (LHC)~\cite{Holo-EWSB1,Holo-EWSB2,CET,piai2}.
Holographic technicolor models are
closely related to extra-dimensional models of electroweak symmetry breaking that
have been studied in much detail recently~\cite{EWSB-models}, but their philosophy is
different. In the AdS/CFT approach, the form of the 5D holographic theory is suggested
by the mechanisms of chiral symmetry breaking and confinement as they appear in string theory
constructions of models similar to QCD.

One benefit of holographic models rooted in D-brane configurations of
string theory, as opposed to in the phenomenological models of
Refs.~\cite{Holo-EWSB1,Holo-EWSB2,CET,piai2},
is that the large-$N$ and curvature corrections, for example, are well
defined and can, in principle, be calculated or estimated systematically.
Furthermore, the gauge theories described by low-energy fluctuations of D-brane
configurations are known, so that both sides of the holographic duality
are well understood. The stringy model of this sort most similar to the
light quark sector of QCD is the D4-D8-\Dbar model of Sakai and
Sugimoto~\cite{SS1,SS2}, which we review in Section~\ref{sec:SS}.  The
Sakai-Sugimoto model is nonsupersymmetric, contains $N_f$ flavors of
massless quarks transforming in the fundamental representation of the SU$(N)$ gauge group,
and is confining with non-Abelian chiral symmetry breaking. (A related model based on
D6-branes rather than D8-branes does not have this last feature~\cite{Mateos}.)
At large $N$ and large 't Hooft coupling $(g^2N)$ with $N_f\ll N$,
the spectrum of resonances, their decay widths and couplings can be reliably
calculated~\cite{SS1,SS2}.

Minimal technicolor models are essentially scaled up versions of QCD, in which an
SU(2)$\times$U(1) subgroup of the chiral symmetry is gauged.  Chiral symmetry
breaking due to the strong technicolor interactions then translates into electroweak
symmetry breaking.  A nice feature of technicolor models is that they avoid the hierarchy
problem associated with scalar Higgs fields.  To produce fermion masses, however, the technicolor
sector must be extended, and it is challenging to accommodate a heavy top quark in such
models while avoiding large flavor-changing neutral current effects.  Also disappointing
is that corrections to precision electroweak observables are estimated to be generically
too large to be consistent with LEP data~\cite{PT}.  Walking technicolor
models, with couplings that run more slowly than in QCD, provide one
approach to solving these problems (see, for example, Refs.~\cite{walking}).

If the electroweak subgroup of the chiral symmetry in the Sakai-Sugimoto model
is gauged, then the model becomes a technicolor-like model of electroweak symmetry breaking.
The Sakai-Sugimoto model has, in addition to $N$, $N_f$ and the 't Hooft coupling, a free
parameter $U_0$, not present in QCD, that fixes the D8-\Dbar brane configuration.  (The definition
of $U_0$ is given in Section~\ref{sec:SS}.) We can therefore hope that for some choice of $U_0$
the model will be consistent with precision electroweak constraints.  Indeed, in phenomenological
holographic technicolor models, additional parameters not present in QCD allow for corrections
to precision electroweak observables that render them consistent with current
bounds~\cite{Holo-EWSB1,Holo-EWSB2,CET,piai,agashe}. The Sakai-Sugimoto model also includes a larger group
of symmetries (a global SO(5) symmetry) and states (Kaluza-Klein modes around a circle of size
comparable to the confining scale) not present in QCD.  The existence of these additional states
may be considered a prediction of the model.  We find it interesting that in
calculable string theory models with chiral symmetry breaking it seems to be difficult to
separate additional physics from the confining scale.  In our adaptation of the Sakai-Sugimoto
model to EWSB, one expects technicolor-like resonances to be discovered together with
Kaluza-Klein modes of a large extra dimension.

The most serious problem for old-fashioned technicolor models is the generically large value of
the Peskin-Takeuchi $S$-parameter, which parameterizes a class of corrections to precision
electroweak observables~\cite{PT}. In Section~\ref{sec:Modes}, we calculate several of the lightest
vector and axial vector resonance masses, their decay constants, and their contribution
to the $S$-parameter, as a function of $U_0$ (the free parameter in the model) for fixed
$g^2N$ and $N$.  We find that the contribution of the lightest resonances to $S$ is
negative, which would seem to make such models promising candidates for a theory of EWSB. However,
we also sum over all modes via a holographic sum rule and demonstrate
a surprising non-decoupling of heavy resonances.  The non-decoupling of contributions to the $S$-parameter
is a consequence of the rapid growth of the decay constants as a function of resonance number, which
offsets the suppression from the increase in resonance masses.
Including the effects of the entire tower of resonances, we find that the $S$-parameter in the model
is generally large and positive.

The decoupling of the heavier resonances can be recovered if we introduce an artificial
ultraviolet (UV) regulator that brings the D8-brane boundary in from infinity.
Although only the first few resonances are
required in this limit to accurately compute $S$, we then find that their contribution typically becomes
positive and large.  For regulator scales lower than the weak scale, we show that it is possible to obtain positive
values of $S$ consistent with current bounds.  We discuss the phenomenological implications of these
results, and other approaches to decreasing the $S$ parameter, in Section~\ref{sec:Conclusions}.

\section{The D4-D8-$\mathbf{\overline{D8}}$ System}\label{sec:SS}
The D4-D8-\Dbar system of Sakai and Sugimoto \cite{SS1,SS2} is similar to QCD
in many ways.  If there are $N$ D4 branes and $N_f$ sets of D8 and \Dbar
branes, then the D-brane configuration describes an SU($N$) gauge theory coupled to
$N_f$ flavors of fermions in the fundamental representation, which we will refer to as
either quarks or techniquarks depending on the context.  The D4 branes are wrapped on a
circle; antiperiodic boundary conditions for the fermions break supersymmetry and lift the masses of
the extraneous (adjoint) fermions. The quarks experience chiral symmetry breaking and confinement, as
is qualitatively understood by properties of the D8-brane configuration and the background geometry.
Properties of the hadronic bound states (masses, decay rates and couplings)
in this system are calculable, as we review here.

\begin{figure}
\begin{center}
$\begin{array}{ccc}
%\multicolumn{1}{l}
%{\mbox{\bf }} &
%   \multicolumn{1}{l}{\mbox{\bf }} \\ [-0.53cm]
\epsfxsize=3in
\epsffile{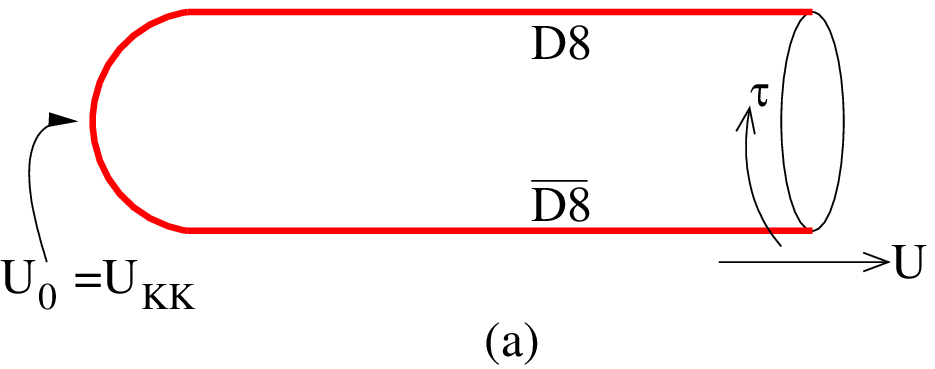} &\ \ \  &
    \epsfxsize=3in
    \epsffile{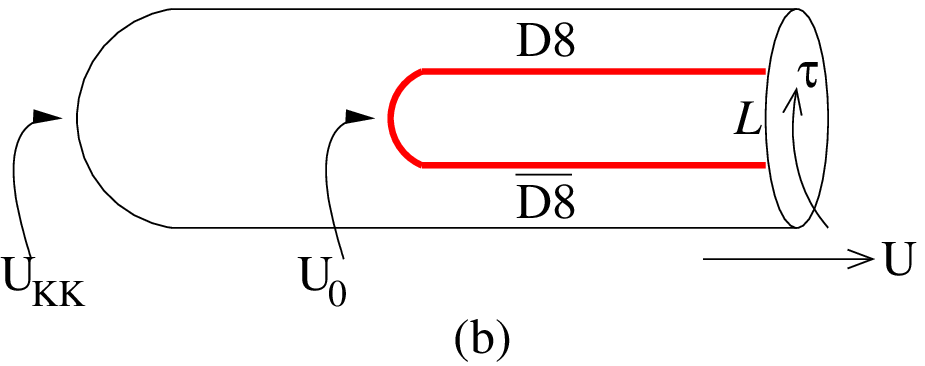} % \\ [0.4cm]
%\mbox{(a)} & \mbox{(b)}
\end{array}$
\end{center}
\caption{\label{fig:U-tube} D8 and \Dbar branes in the D4-brane geometry.
a) Antipodal D8-\Dbar configuration; b) generic D8-\Dbar configuration.}
\end{figure}

The D4 brane geometry, in the notation of \cite{SS1} (except for the
signature of the metric), is given by,
\begin{equation}
ds^2=\left(\frac{U}{R}\right)^{3/2}\left(\eta_{\mu\nu}dx^\mu dx^\nu-f(U)
d\tau^2\right)- \left(\frac{R}{U}\right)^{3/2}\left(\frac{dU^2}{f(U)}+
U^2\,d\Omega_4^2\right),
\label{eq:themetric}
\end{equation}
where
\begin{equation}
f(U)=1-\frac{U_{KK}^3}{U^3} \,\,\,\, \mbox{ and } \,\,\,\, \quad R^3=\pi g_sN l_s^3.
\end{equation}
The D4 brane extends in the $x^\mu, \mu=0,1,2,3$ and $\tau$ directions, where
the $\tau$ direction is compactified on a circle. In the remaining dimensions, the metric is spherically
symmetric: $U$ is the radial coordinate and $d\Omega_4^2$ is the metric of
the unit 4-sphere.  The scale $U_{KK}$ is a free parameter, but to avoid a singularity at
$U=U_{KK}$ the variable $\tau$ is periodic with period $4\pi R^{3/2}/(3 U_{KK}^{1/2})$.  We
will refer to the geometry projected onto the $\tau$ and $U$ directions
as the $U$-tube, illustrated in Fig.~\ref{fig:U-tube}.  The D8
and \Dbar branes extend in the $x^\mu$ and 4-sphere directions, and follow a minimal energy
trajectory $U(\tau)$ on the $U$-tube. Physical results will depend on
$R$ and $U_{KK}$ in the combination,
\begin{equation}
\frac{R^3}{U_{KK}}\equiv \frac{9}{4}M_{KK}^{-2},
\end{equation}
so that there is a single dimensionful scale, $M_{KK}$, governing the dynamics of the model.  The
boundary of the spacetime is at $U=\infty$, which is topologically $S^1\times S^4\times M^4$, where
$M^4$ is 3+1 dimensional Minkowski space (with the Lorentz invariance of the 4D theory).
The dilaton appears in the Dirac-Born-Infeld (DBI) action governing the dynamics of the D8 branes;
in the D4 brane background the dilaton profile is,
\begin{equation}
e^\phi=g_s\left(\frac{U}{R}\right)^{3/4}.
\end{equation}
The RR four-form is also turned on in the D4-brane background, but we will not need its
profile in what follows.

Confinement in the model is related to the termination of the geometry at $U_{KK}$, {\em i.e.}
the restriction $U\geq U_{KK}$.  One way to understand confinement in this context is to
consider a string stretched from the boundary at one point in $M^4$ to another as a
function of separation of the string endpoints. The minimum energy string extends
away from the boundary, and for large enough separation between string endpoints,
will stretch all the way to $U_{KK}$.  For still larger separation, the string
stretches from infinity along the $U$-tube to $U_{KK}$, moves along $M^4$, and returns
to $U=\infty$.  The motion along $M^4$ at roughly fixed $U\approx U_{KK}$
gives a contribution to the energy of the string proportional to the separation
of the endpoints along $M^4$.  This linear potential is a signature of confinement.

One striking difference between QCD and the Sakai-Sugimoto model is that the latter
predicts additional Kaluza-Klein (KK) modes associated with the compact dimensions.
Notably, $M_{KK}$ sets both the confinement scale and the mass scale of the KK modes;
above the confining scale, the additional dimension parameterized by the coordinate $\tau$
becomes apparent.  While it is common to ignore the Kaluza-Klein modes, and assume that
only the 4D model of the lowest modes is realistic, we consider the existence of
these Kaluza-Klein excitations as a prediction of holographic EWSB.  Thus, we expect a tower
of heavy gauge bosons to appear in addition to technicolor-like resonances. The D4-D8-\Dbar
configuration also has an SO(5) symmetry from the 4-sphere directions, which becomes
an SO(5) global symmetry in the field theory, leading to additional KK modes.

\begin{figure}
\includegraphics[scale=.6]{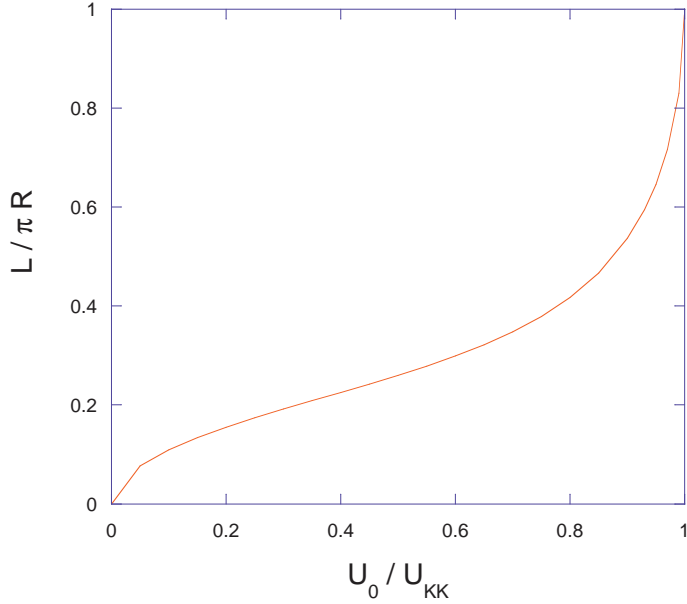}
\caption{\label{fig:LpiR} The asymptotic distance between D8 and \Dbar branes
along the $\tau$ circle as $U\rightarrow\infty$ as a function of
$U_{0}/U_{KK}$.}
\end{figure}

In order to decouple string loop ($g_s$) and gravity ($\alpha'$) corrections,
$N$ and $g_sN$ are taken to be large.  In the probe brane limit, $N_f\ll N$
(first introduced in the context of the D3-D7 system by Karch and Katz~\cite{KK}),
the D4 branes generate a spacetime geometry in which stable D8 brane configurations
minimize their action. This is the easiest case to study, as the backreaction on the
geometry due to the D8 branes can be ignored. The D8-brane configurations in the
D4-D8-\Dbar system were worked out in Refs.~\cite{SS1} and~\cite{Aharony}.
It was found that the D8 and \Dbar branes (which are otherwise
distinguished by their orientation)
join at some $U_0\geq U_{KK}$.  The D8-\Dbar profile $U(\tau)$ is determined by
minimizing the Dirac-Born-Infeld action with gauge fields on the branes turned off,
\begin{equation}
S_{DBI}=-T\int d^9x\, e^{-\phi} \sqrt{{\rm det}\,g_{MN}},
\end{equation}
where $g_{MN}$ is the induced metric on the D8-branes.  The result of minimizing
this action is~\cite{Aharony},
\begin{equation}
f(U)+\left(\frac{R}{U}\right)^3\frac{U'(\tau)^2}{f(U)}
=\frac{U^8 f(U)^2}{U_0^8f(U_0)} \,\,\,.
\label{eq:D8config}
\end{equation}
For the solution with $U_0=U_{KK}$, the D8-\Dbar branes stretch from antipodal points along
the $\tau$ circle at $U=\infty$  to the tip of the $U$-tube at $U=U_{KK}$, as in
Fig.~\ref{fig:U-tube}a. Generic D8-\Dbar configurations are sketched in
Fig.~\ref{fig:U-tube}b.
For the solutions with $U_0>U_{KK}$ the asymptotic distance along
the $\tau$ circle between the D8 and \Dbar branes as $U\rightarrow\infty$ is found to be~\cite{Aharony}
\begin{equation}
L=2R^{3/2}\int_{U_0}^\infty dU\,\frac{1}{f(U)U^{3/2}\sqrt{\frac{f(U)U^8}{
f(U_0)U_0^8}-1}}.
\end{equation}
The value of $L/\pi R$ is plotted as a function of $U_{0}/U_{KK}$ in
Fig.~\ref{fig:LpiR}.

The induced metric on the D8-\Dbar configuration is obtained by eliminating $d\tau^2$
in favor of $dU^2$ in Eq.~(\ref{eq:themetric}) using Eq.~(\ref{eq:D8config}):
\begin{equation}
ds^2=\left(\frac{U}{R}\right)^{3/2}\,\eta_{\mu\nu}dx^\mu dx^\nu-
\left(\frac{R}{U}\right)^{3/2}\,U^2\,d\Omega_4^2-
\left(\frac{R}{U}\right)^{3/2}\left[\frac{1}{f(U)}+\left(\frac{U}{R}\right)^3
\frac{f(U)}{U'(\tau)^2}\right]\,dU^2. \label{eq:induced}\end{equation}
The determinant of the induced metric is then
\begin{equation}
\sqrt{\det\, g}=\frac{U^{29/4}R^{3/4}\sqrt{\det\,g_{\Omega_4}}}{\left(
U^8f(U)-U_0^8f(U_0)\right)^{1/2}}.
\label{eq:detinduced}
\end{equation}
The connection of the branes at $U_0$ signals the breaking of the
SU$(N_f)\times$SU$(N_f)$ chiral symmetry of the 4+1 dimensional theory
(or correspondingly the gauge invariance on the D8 branes) to an SU$(N_f)$
subgroup.  Quarks in this theory are massless because the D8 branes
and D4 branes cannot be separated: the D8 branes are codimension 1 and the
D4 branes extend in the transverse direction.  Chiral symmetry breaking
is associated with the formation of a quark condensate, although it is
difficult to ascertain its exact value as a function of $U_0$.  (If the quark mass
were nonvanishing, one could vary the action with respect to it to obtain
the condensate, as in Ref.~\cite{Mateos}.  For nonvanishing quark masses
in D8-brane scenarios, see Ref.~\cite{quarkmass}.)  On the other hand,
the technipion decay constant can be estimated holographically, as we
describe in the next section, and is fixed to $f_\pi=246$ GeV in order to
reproduce the spectrum of electroweak gauge bosons (or to $f_\pi=92$ MeV
to reproduce the pion decay constant in QCD). In QCD, the confinement and
chiral symmetry breaking scales are related to one another, as is the case in
the D-brane construction.  However, in the D4-D8-\Dbar model there is an additional
parameter, $U_0/U_{KK}$, that does not correspond to any parameter in QCD.
The ratio of the chiral symmetry breaking scale to the confinement scale
varies as a function of the free parameter, as noticed in Ref.~\cite{Harvey}.

\section{Holographic Technicolor from the D4-D8-$\mathbf{\overline{D8}}$
System}\label{sec:Modes}
In technicolor models, a new set of asymptotically free gauge interactions
are postulated under which the Standard Model particles are neutral.
Fermions charged under both the technicolor and electroweak interactions
condense due to strong technicolor dynamics and break the electroweak
gauge group to the U(1) gauge invariance of electromagnetism.  Just as hadrons
appear as bound states of confined quarks and gluons, technihadrons are
predicted to exist as bound states of techniquarks and technigluons.
The properties of technihadrons are difficult to calculate for the same reason that
the properties of hadrons in QCD are difficult to calculate: the interactions are strong
at the low energies of interest.  However, in holographic models it is straightforward
to estimate masses, decay rates and couplings of hadronic resonances. In QCD, the
chiral symmetry of the light quarks is broken to a diagonal subgroup, yielding a
spectrum of Goldstone bosons that we refer to generically as pions.  In technicolor models,
an SU(2)$\times$U(1) subgroup of the chiral symmetry is gauged and identified with the
electroweak gauge group.  With two techniquark flavors, the would-be pions are eaten by
the $W$ and $Z$ boson and do not appear as physical states in the spectrum.  However,
techni-vector ($\rho$) resonances and techni-axial vector ($a_1$) resonances are expected
to be present, and are a generic prediction of technicolor models.

In phenomenological models of EWSB based on the AdS/CFT correspondence,
it has been found that corrections to precision electroweak observables
can be suppressed by adjusting available
parameters~\cite{Holo-EWSB1,Holo-EWSB2,CET,piai}.  There are even
scenarios in which the $S$-parameter is negative \cite{Holo-EWSB2}, although
there is some debate as to whether scenarios leading to negative $S$
can be physically realized \cite{Holo-EWSB1,csaba}.
A disadvantage to the phenomenological approach is that it is not known
whether an ordinary quantum field theory interpretation of those
models is always possible.  On the other hand, models based on actual D-brane
configurations avoid this problem: the appropriate field theory description is
that of low-energy fluctuations on the given D-brane configuration.
Unfortunately, the AdS/CFT correspondence is most powerful as a calculational
tool in limits of the parameters that may not be physically relevant.  In
particular, in the supergravity limit it is assumed that $N$ and $g^2 N$ are
large, so that gravity and stringy effects are decoupled from the field
theory that lives on the D-branes.  As a result of the large-$N$ limit, the
resonances are infinitely narrow in a spectral decomposition of correlation
functions, an outcome that is generic in holographic models if higher dimensional
loop corrections are not taken into account.

One of the most serious problems for technicolor models is their generically
large value of the Peskin-Takeuchi $S$-parameter~\cite{PT}. The $S$ parameter
can be defined in terms of matrix elements of the vector current $J_\mu^{a\,V}$
and axial-vector current $J_\mu^{a\,A}$ two-point functions \cite{PT}:
\begin{eqnarray}
i\,\int d^4x\,e^{iq\cdot x}\,\langle J_\mu^{a\,V}(x)J_\nu^{b\,V}(0)\rangle &=&
\left(-g_{\mu\nu}+\frac{q_\mu q_\nu}{q^2}\right) \,\delta^{ab}\,\Pi_V(-q^2), \nonumber \\
i\,\int d^4x\,e^{iq\cdot x}\,\langle J_\mu^{a\,A}(x)J_\nu^{b\,A}(0)\rangle &=&
\left(-g_{\mu\nu}+\frac{q_\mu q_\nu}{q^2}\right)\, \delta^{ab}\,\Pi_A(-q^2)
\end{eqnarray}
\begin{equation}
S=-4\pi \frac{d}{dQ^2}\left.\left(\Pi_V-\Pi_A\right)
\right|_{Q^2=0} \,\,,
\label{eq:S}
\end{equation}
where $Q^2=-q^2$. From a dispersive representation with delta
function resonances, the current two-point functions can be
expressed in terms of the resonance masses and decay constants (up
to constant local counterterms) yielding
\begin{eqnarray}
\Pi_V(-q^2)&=&\sum_n\frac{g_{Vn}^2 \, q^2}{m_{Vn}^2(-q^2+m_{Vn}^2)} \,\,\, , \nonumber \\
\Pi_A(-q^2)&=&-f_\pi^2+\sum_n\frac{g_{An}^2 \, q^2}{m_{An}^2(-q^2+m_{An}^2)} .
\label{eq:Pi}
\end{eqnarray}
It then follows that the $S$ parameter can be written as a sum over
the vector and axial vector resonances as
\begin{equation}
S=4\pi\sum_n\left(\frac{g_{Vn}^2}{m_{Vn}^4}-\frac{g_{An}^2}{m_{An}^4} \right).
\end{equation}
The spectrum and decay constants of the lightest few hadronic resonances in
the Sakai-Sugimoto model for the configuration $U_{0}=U_{KK}$ in which the D8-branes stretche
between antipodal points on the $\tau$ circle were calculated in Ref.~\cite{SS2}.
The extension of these results to other D8-brane configurations was described in
Ref.~\cite{Aharony}, although the purpose of that paper was to study finite temperature
physics, and numerical solutions for the resonance masses and decay constants were not
computed. The analysis is similar to that for the D4-D6-$\overline{\mbox{D6}}$ system
in Ref.~\cite{Mateos}.

To calculate the decay constants $g_{Vn}$ and $g_{An}$, and masses
$m_{Vn}$ and $m_{An}$, one studies the DBI action on the probe D8-branes,
\begin{equation}
S_{D8}=-T\int_{{\rm D}8+\dbar} d^4x\,dU\,d\Omega_4\,e^{-\phi}
\sqrt{\det\,(g_{MN}+(2\pi\alpha')F_{MN})}\,\,\, ,\end{equation}
where $F_{MN}=F_{MN}^aT^a$ is the gauge field strength on the D8 branes, $\alpha'=l_s^2$,
and $g_{MN}$ is the induced metric, Eq.~(\ref{eq:induced}).  The determinant is over the
Lorentz matrix structure, and traces over the gauge group in the expansion of the
determinant are to be understood.
There is a Chern-Simons term in the D8-brane action which we consistently ignore,
since we restrict our attention to terms only quadratic in the gauge fields.
The integral is over $U\in(U_0,\infty)$ twice: once
over the D8 brane segment and once over the \Dbar brane segment.
We only study configurations constant along the $S^4$ and such that $F_{MN}=0$
for $M,N \neq 0,1,2,3,U$. From a 5D perspective, the remaining modes are
related to these by the SO(5) symmetry of the brane configuration, plus
a set of scalar fields from the decomposition of the gauge fields around the
4-sphere.  Expanding the action to quadratic order in $F_{MN}$,
\begin{eqnarray}
S_{D8} &\approx&
-\frac{3}{2}\Ttilde(2\pi\alpha')^2R^3 U_{KK}^{-1/2} \cdot \nonumber \\
&& \int d^4x\,dU\,\Tr\left[\frac{1}{2}F_{\mu\nu}
F^{\mu\nu}\,U^{-1/2}\gamma(U)^{1/2}+ F_{\mu U}F^{\mu U}\,U^{5/2}
R^{-3}\gamma(U)^{-1/2}\right],
\label{eq:quad-action}
\end{eqnarray}
where $\Ttilde=\frac{2}{3} R^{3/2}U_{KK}^{1/2}T V_4 g_s^{-1}$, $V_4=8\pi^2/3$
is the area of the unit 4-sphere, and contractions of indices are with respect
to the 5D Minkowski metric.  We have defined the function $\gamma(U)$ as in
Ref.~\cite{Aharony}:
\begin{equation}
\gamma(U)=\frac{U^8}{U^8 f(U)-U_0^8f(U_0)}.
\end{equation}

Expanding $F_{MN}$ in modes, the vector modes are symmetric upon reflection about
$U=U_0$ while the axial vector modes are antisymmetric. These modes are identified with
the tower of vector and axial vector mesons in the 4+1 dimensional gauge theory,
with the $\tau$ circle dimension suppressed.  The normalizable modes satisfy Dirichlet
boundary conditions: $A_\mu(x,\infty)=0$ on both the D8 and \Dbar branches of the
$U$-segment $U\in(U_0,\infty)$.  The symmetric solutions also satisfy
$\partial_U A_\mu|_{U=U_0}=0$, while the antisymmetric modes satisfy $A_\mu(x,U_0)=0$.

Note that in the $U$ coordinate system one must consider two branches of
$U\in(U_0,\infty)$, the D8-brane branch and the \Dbar-brane branch.  The branes meet
at $U_0$ so we sometimes refer simply to the D8-brane configuration.
To simplify the discussion of boundary conditions we will sometimes find it convenient
to change coordinates, for example to $s$ defined by $s^2=U-U_0$, $s\in(-\infty,\infty)$,
which covers both branches of the D8-\Dbar configuration.  Since the
equations of motion are more cumbersome in such coordinate systems, unless $U_0=
U_{KK}$, we will use the $U$ coordinate in most of what follows.

\begin{figure}
\includegraphics[scale=.5]{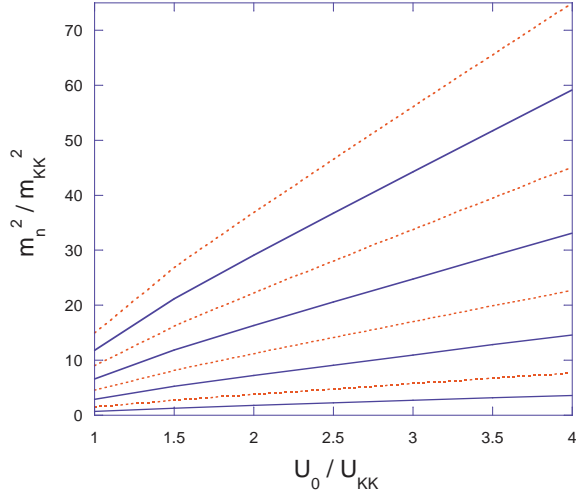}
\caption{\label{fig:masses} Masses of the lightest four vector (solid lines)
and axial vector (dotted lines) resonances as a function of $U_0/U_{KK}$.}
\end{figure}

We will work in a gauge in which $A_U(x,u)=0$.
As in Ref.~\cite{SS2},
we expand the vector field in normalizable modes satisfying Dirichlet boundary conditions
$A_\mu^n(x,\infty)$=0:\footnote{Since we are interested in the case $N_f=2$
we should be careful not to fall prey to the Witten anomaly.  To be precise,
the boundary condition is that the gauge fields are pure gauge at
infinity, and such configurations
fall into one of two classes (because $\pi_4(SU(2))=\mathbb{Z}_2$).
This subtlety will not affect any of our results, and we simply note that
we must assume the techniquarks
transform under an even-dimensional representation of the technicolor group
to have a well-defined theory.}
\begin{equation}
A_\mu(x,U)=\sum_n \left(V_\mu^n(x)\psi_{Vn}(U)+
A_\mu^n(x)\psi_{An}(U)\right).
\end{equation}
The $\psi_{Vn}(U)$ are the symmetric modes, and $\psi_{An}(U)$ are
antisymmetric. The equations of motion are
\begin{eqnarray}
&-U^{1/2}\gamma^{-1/2}\partial_U\left(U^{5/2}\gamma^{-1/2}\partial_U\psi_{Vn}
\right)=R^3 m_{Vn}^2\psi_{Vn},& \nonumber \\
&\partial_\mu F_V^{\mu\nu}(x) = -m_{Vn}^2 V_n^\nu(x),&
\label{eq:EOM}
\end{eqnarray}
and similarly for the axial vector modes with $V\rightarrow A$.  The spectrum
alternates between symmetric and antisymmetric modes, so that
the vector and axial vector masses alternate, as in Fig.~\ref{fig:masses}.
A source for the SU(2)$_L$ or SU(2)$_R$ current corresponds to a non-normalizable
solution to the equation of motion for $F_{\mu\nu}$ localized near either the D8 or
\Dbar brane boundary, respectively. Similarly, a source for SU(2)$_V$ corresponds
to a non-normalizable solution symmetric around $U=U_0$, while a source for
SU(2)$_A$ corresponds to an anti-symmetric configuration.  We will call the
symmetric non-normalizable solutions, Fourier transformed in the
coordinates $x^\mu$,  $A_\mu(q,U)={\cal V}_\mu(q)\psi_V^0(q^2,U)$,
where $\psi_V^0(q^2,\infty)=1$ on both branches of $U$. We call the
antisymmetric solutions $A_\mu(q,U)={\cal A}_\mu(q)\psi_A^0(q^2,U)$, where
$\psi_A^0(q^2,\infty)=1$ on the D8 brane branch and $-1$ on the \Dbar branch.
The gauge transformation which sets $A_U(q,U)=0$ reappears in the sources
${\cal V_\mu}$ and ${\cal A_\mu}$ and gives rise to the pion kinetic
terms and couplings, as discussed in \cite{SS2}.
For our purposes it will be good enough to set these
pion degrees of freedom to zero.  They will get eaten after gauging of the
electroweak subgroup of the chiral symmetry, in any case.
The bulk-to-boundary propagators $\psi_V^0(q^2,U)$ and $\psi_A^0(q^2,U)$ satisfy
the equation of motion (\ref{eq:EOM}) with $m_{Vn}^2$ replaced by $q^2$.  A
general decomposition of the gauge fields into normalizable modes in
addition to a source
for the SU(2)$\times$SU(2) current is then
\begin{equation}
A_\mu(q,U)=\sum_n\left(V_\mu^n(q)\psi_{Vn}(U)+A_\mu^n(q)\psi_{An}(U)\right)
+{\cal V}_\mu(q)\psi_V^0(q^2,U)+{\cal A}_\mu(q)\psi_A^0(q^2,U).
\end{equation}

Up to gauge fixing terms, the decomposition of the action, Eq.~(\ref{eq:quad-action}),
to quadratic order in the gauge fields is given by \pagebreak
\begin{eqnarray}
S_{D8}&=& -\frac{3}{2}\Ttilde(2\pi\alpha')^2R^3U_{KK}^{-1/2} \nonumber \\
&& \,{\rm Tr}\int_{{\rm D8}+\dbar} d^4q\,dU
\left\{\frac12 \left[|F_{\mu\nu}^{V0}(q)|^2(\psi_{V}^0(q^2,U))^2+
|F_{\mu\nu}^{A0}(q)|^2(\psi_{A}^0(q^2,U))^2\right.\right. \nonumber
\\
&&+\sum_n\left.\left(|F_{\mu\nu}^{Vn}(q)|^2\psi_{Vn}^2+
|F_{\mu\nu}^{An}(q)|^2\psi_{An}^2\right)\right]
\frac{\gamma^{1/2}}{U^{1/2}} \nonumber
\\
&&-\left[(\partial_U\psi_{V}^0)^2\,|V_\mu^0(q)|^2+
(\partial_U\psi_{A}^0)^2\,|A_\mu^0(q)|^2 \right.\nonumber \\
&&+\sum_n \left.\left(
(\partial_U\psi_{Vn})^2\,|V_\mu^n(q)|^2+
(\partial_U\psi_{An})^2\,|A_\mu^n(q)|^2\right)\right]
\frac{U^{5/2}}{\gamma^{1/2}} \nonumber \\
&&+\left.\left(F_{\mu\nu}^{V0}(q)F^{\mu\nu}_{Vn}(-q)\, \psi_{Vn}\psi_V^0+
F_{\mu\nu}^{A0}(q)F^{\mu\nu}_{An}(-q)\, \psi_{An}\psi_A^0\right)\frac{\gamma^{
1/2}}{U^{1/2}}\right\}. \end{eqnarray}
If the SU$(N_f)$ generators are normalized so that Tr\,$T^aT^b=\delta^{ab}/2$,
then the gauge fields are canonically normalized if:
\begin{eqnarray}
3\Ttilde(2\pi\alpha')^2R^3U_{KK}^{-1/2}\,\int_{U_0}^\infty dU
\psi_{Vn}\psi_{Vm}\gamma^{1/2}U^{-1/2} = \delta_{mn}, \label{eq:cnormv} \\
3\Ttilde(2\pi\alpha')^2R^3U_{KK}^{-1/2}\,\int_{U_0}^\infty dU
\psi_{An}\psi_{Am}\gamma^{1/2}U^{-1/2} = \delta_{mn}.
\end{eqnarray}
It is natural to define, as in Ref.~\cite{SS2},
\begin{equation}
\kappa\equiv\Ttilde (2\pi\alpha')^2R^3=\frac{g^2N^2}{108\pi^3},
\end{equation}
where we have used the relations \cite{SS1}:
\begin{equation}
R^3=\frac{g^2N^2 l_s^2}{2M_{KK}},\quad U_{KK}=\frac{2}{9}g^2NM_{KK}l_s^2,
\quad g_s=\frac{g^2}{2\pi M_{KK}l_s}.
\end{equation}
Using the equations of motion Eq.~(\ref{eq:EOM}) we can write the action in
terms of the canonically normalized modes as
\begin{eqnarray}
S_{D8}&=&-{\rm Tr}\,\int d^4x\,\sum_n\left[\frac12 (F_{\mu\nu}^{Vn})^2+\frac12
(F_{\mu\nu}^{An})^2 -m_{Vn}^2\,(V_\mu^n)^2 -m_{An}^2\,(A_\mu^n)^2 \nonumber
\right.\\
&&+\left.\left(a_{Vn}F_{\mu\nu}^{V0}F_{Vn}^{\mu\nu}+
a_{An}F_{\mu\nu}^{A0}F_{An}^{\mu\nu}\right) \right] +S_{{\rm source}},
 \end{eqnarray}
where
\begin{eqnarray}
a_{Vn}&=&-
\left.3\kappa (m_{Vn}^2R^3)^{-1}U^{5/2}U_{KK}^{-1/2}\gamma^{-1/2}\partial_U\psi_{Vn}\right|
_{U_{{\rm D}8}=\infty}, \nonumber \\
a_{An}&=&-
\left.3\kappa (m_{An}^2R^3)^{-1}U^{5/2}U_{KK}^{-1/2}\gamma^{-1/2}\partial_U\psi_{An}\right|
_{U_{{\rm D}8}=\infty}.
\label{eq:aVaA}\end{eqnarray}
$S_{{\rm source}}$ is the kinetic term for the sources ${\cal V}$ and
${\cal A}$, and is given by,
\begin{equation}
S_{{\rm source}}=
-{\rm Tr}\,\int d^4q\,q^2\,\left[a_{V0}\,|V_\mu^0(q)|^2+
a_{A0}\,|A_\mu(q)|^2\right]. \label{eq:Ssource}\end{equation}
The constants $a_{V0}$ and $a_{A0}$ are given by,
\begin{eqnarray}
a_{V0}&=&-
\left.3\kappa (q^2R^3)^{-1}U^{5/2}U_{KK}^{-1/2}\gamma^{-1/2}\partial_U\psi_V^0
(q^2,U)\right|
_{U_{{\rm D}8}=\infty}, \nonumber \\
a_{A0}&=&-
\left.3\kappa (q^2R^3)^{-1}U^{5/2}U_{KK}^{-1/2}\gamma^{-1/2}\partial_U\psi_A^0
(q^2,U)\right|
_{U_{{\rm D}8}=\infty}.
\label{eq:aV0aA0}\end{eqnarray}

We have used the equations of motion and boundary conditions on $\psi_{Vn}, \psi_{An}$,
$\psi^0_{V}$, and $\psi^0_{A}$ to eliminate vanishing surface terms, {\em e.g.},
\begin{eqnarray}
&&\int_{{\rm D}8+\dbar} dU\,U^{-1/2}\gamma^{1/2}\psi_{Vn}(U)\,\psi_V^0(q^2,U) \nonumber \\
&=&-\int_{{\rm D}8+\dbar} dU\, (m_{Vn}^2R^3)^{-1}\partial_U\left(
U^{5/2}\gamma^{-1/2}\,\partial_U\psi_{Vn}\right)\psi_V^0 \nonumber \\
&=&-\left.2(m_{Vn}^2R^3)^{-1}U^{5/2}\gamma^{-1/2}\partial_U\psi_{Vn}\right|
_{U_{{\rm D}8}=\infty},
\end{eqnarray}
and similarly for $V\rightarrow A$.

\begin{figure}
\includegraphics[scale=.6]{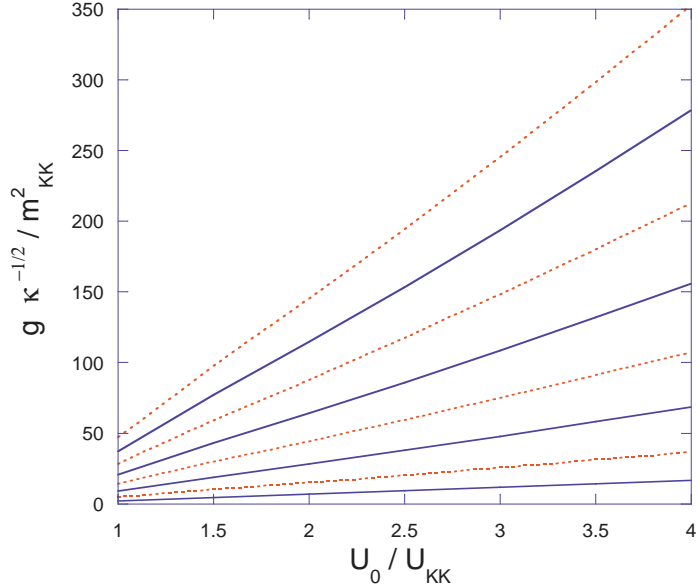}
\caption{\label{fig:g} Decay constants of the lightest four vector (solid
lines) and
axial vector (dotted lines) resonances as a function of $U_0/U_{KK}$.}
\end{figure}

\begin{figure}
\includegraphics[scale=.6]{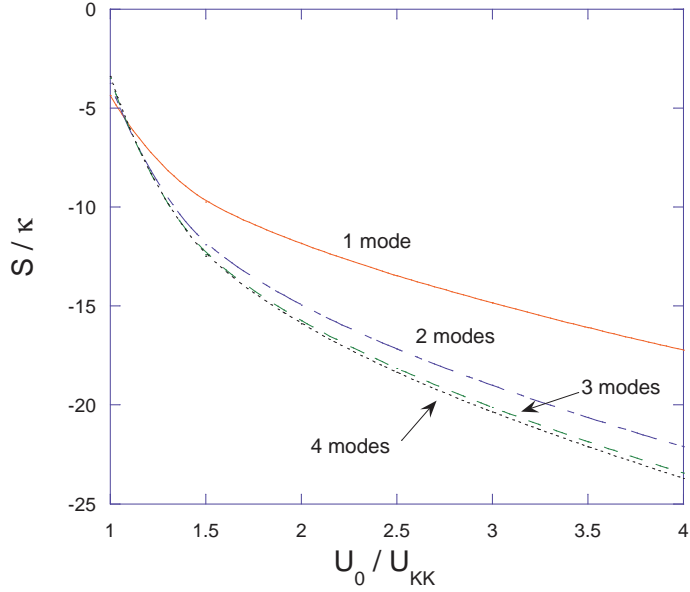}
\caption{\label{fig:S4} Partial sums of the lightest four vector and axial vector
resonance contributions to the $S$ parameter, as a function of $U_0/U_{KK}$.}
\end{figure}

Diagonalizing the kinetic terms, we obtain the action,
\begin{eqnarray}
S_{D8}&=&-
{\rm Tr}\,\int d^4x\,\sum_n\left[\frac12 (\widetilde{F}_{\mu\nu}^{Vn})^2
-m_{Vn}^2\,(\widetilde{V}_\mu^n-a_{Vn}{\cal V}
_\mu)^2 \right.\nonumber \\
&&+\left.\frac12
(\widetilde{F}_{\mu\nu}^{An})^2 -m_{An}^2\,(\widetilde{A}_\mu^n-a_{An}{\cal A}
_\mu)^2
\right] +\widetilde{S}_{{\rm source}}, \end{eqnarray}
where
\begin{equation}
\widetilde{V}^n_\mu=V^n_\mu+a_{Vn}{\cal V}_\mu,
\end{equation}
$\widetilde{F}_{\mu\nu}^{Vn}$ is the gauge kinetic term for $\widetilde{V}^n
_\mu$,
and similarly for $V\rightarrow A$.  The kinetic term for the source
picks up an additional contribution in this process.
From the coupling between the sources and the modes we identify the decay
constants,
\begin{eqnarray}
g_{Vn}&=&m_{Vn}^2 a_{Vn} \nonumber \\
&=&-\left.3\kappa R^{-3}U^{5/2}U_{KK}^{-1/2}\gamma^{-1/2}\partial_U\psi_{Vn}\right|
_{U_{{\rm D}8}=\infty}, \nonumber \\
g_{An}&=&m_{An}^2 a_{An} \nonumber \\
&=&-\left.3\kappa R^{-3}U^{5/2}U_{KK}^{-1/2}\gamma^{-1/2}\partial_U\psi_{An}\right|
_{U_{{\rm D}8}=\infty}. \label{eq:thedcs}
\end{eqnarray}
The decay constants grow with $U_0$ as the masses do, as demonstrated in
Fig.~\ref{fig:g}.
The spectrum of the lightest four vector and axial vector resonances, and their
decay constants, were given in Ref.~\cite{SS2} for the antipodal D8-\Dbar
configuration $U_0=U_{KK}$.  The contribution of the first four pairs of resonances
to the $S$ parameter Eq.~(\ref{eq:S}) using the results of Ref.~\cite{SS2} is
$S_4\approx -0.4\kappa \approx -0.006$, where in the last approximation we
set $g^2N=4\pi$ and $N=4$.  The lightest pair of resonances contributes negatively to
the $S$-parameter. In fact, the three-digit precision to which the results in
Ref.~\cite{SS2} are quoted is not good enough for this calculation because of the propagation
of errors in summing over the small differences between the large contributions
of the vector and axial vector modes to $S$.  We find by increasing the precision of
our numerical calculation that $S_4\approx -3\kappa \approx -.05$, larger
by an order of magnitude than our estimate from Sakai and Sugimoto's results.  It is
clear that the contributions from individual sets of modes to the $S$ parameter are sensitive
to the details of the model. Increasing the parameter $U_0/U_{KK}$ makes the contribution
of the lightest modes to the $S$-parameter even more negative, as seen in Fig.~\ref{fig:S4}.

On the other hand, we can also calculate the $S$-parameter exactly (in the
large $N$ limit) by using the bulk-to-boundary propagators $\psi_V^0$ and $\psi_A^0$,
analogous to the derivation of the sum rules in Ref.~\cite{sumrules}.   It is
convenient to temporarily employ a coordinate $s$ that spans both the D8 and \Dbar branches,
\begin{equation}
s^2 = U-U_0 \,\,\, ,
\end{equation}
where $s>0$ ($s<0$) corresponds to the D8 brane (\Dbar brane). In momentum space, the
vector bulk-to-boundary propagator $\psi_V^0(q^2,s)$ satisfies
\begin{equation}
\left[R^3 q^2 U^{-1/2} \gamma^{1/2} + \frac{1}{2 s} \partial_s (\frac{1}{2 s}
U^{5/2} \gamma^{-1/2}\partial_s)\right] \psi_V^0(q^2,s) = 0  \,\,\, ,
\end{equation}
with $\psi_V^0(q^2,s)=1$ at $s=\pm\infty$.  We aim to relate this to the following Green's
function $G(s,s',q^2)$ for the equation of motion operator
\begin{equation}
\left[R^3 q^2 U^{-1/2} \gamma^{1/2} + \frac{1}{2 s}  \partial_s (\frac{1}{2 s} U^{5/2} \gamma^{-1/2}\partial_s)
\right] G(s,s',q^2) = \frac{1}{|2s|} \delta(s-s')  \,\,\, ,
\label{eq:greendef}
\end{equation}
where we assume Dirichlet boundary conditions $G(\pm\infty,s',q^2)=0$.  It follows from
the normalization condition for the modes, Eq.~(\ref{eq:cnormv}), that we may rewrite the delta
function in Eq.~(\ref{eq:greendef}) using the completeness relation
\begin{equation}
\frac{3}{2} U_{KK}^{-1/2} \kappa \, U^{-1/2} \gamma^{1/2}\, \sum_n \psi_{n}(s)
\psi_{n}(s') = \frac{1}{|2s|} \delta(s-s') \,\,\,.
\end{equation}
If we then apply the ansatz
\begin{equation}
G(s,s',q^2)=\sum_{nm} c_{nm} \, \psi_{n}(s)\, \psi_{m}(s') \,\,\,,
\end{equation}
one can extract the coefficients $c_{nm}$ from Eq.~(\ref{eq:greendef}) by application of
the equations of motion.   It follows that $G(s,s',q^2)$ may be expressed as a sum over modes,
\begin{equation}
G(s,s',q^2) = \frac{3}{2} \frac{U_{KK}^{-1/2} \kappa}{R^3} \sum_n \frac{\psi_{n}(s)\psi_{n}(s')}
{q^2-m_{n}^2} \,\,\,.
\end{equation}
For definiteness, let us consider the relationship between $G(s,s',q^2)$ and the bulk-to-boundary
propagator of the vector modes $\psi_V^0(q^2,s)$.  We begin with the identity
\begin{equation}
\frac{s'}{|s'|} \psi_V^0(q^2,s') = \int_{-\infty}^{\infty} ds \, \delta(s-s')\,
\frac{s}{|s|} \psi_V^0(q^2,s) \,\,\,.
\label{eq:dident}
\end{equation}
Replacing the delta function using Eq.~(\ref{eq:greendef}), one may integrate by parts twice
to show that the right-hand-side of Eq.~(\ref{eq:dident}) is a surface term
\begin{equation}
\frac{s'}{|s'|} \psi_V^0(q^2,s') = \frac{1}{s} \, U^{5/2} \gamma^{-1/2}\, \partial_s G_V(s,s',q^2)|^{s=\infty} \,,
\end{equation}
where we have defined $G_V(s,s',q^2)=[G(s,s',q^2)+G(-s,s',q^2)]/2$.  In terms of the mode
decomposition, $G_V(s,s',q^2)$ contains only those modes that are even under reflection
about $s=0$ ({\em i.e.}, $U=U_0$), which we called $\psi_{Vn}(s)$ earlier:
\begin{equation}
\frac{s'}{|s'|} \psi_V^0(q^2,s') = 2 \, U^{5/2} \gamma^{-1/2} \, \left(\frac{3}{2}
\frac{U_{KK}^{-1/2} \kappa}{R^3} \right) \sum_n\left.
\frac{\frac{1}{2s} \partial_s \psi_{Vn}(s)\psi_{Vn}(s')}{q^2-m_{Vn}^2}
\right|^{s=\infty} \,\,\,.
\label{eq:intres1}
\end{equation}
The form of this result is suggestive: $ U^{5/2}
\gamma^{-1/2} \frac{1}{2 s} \partial_s \psi_{Vn}(s)|^{s=\infty}$ is precisely
the same  as the quantity $U^{5/2} \gamma^{-1/2} \partial_U \psi_{Vn}(U)|_{U_{D8}=\infty}$ that
appeared in our holographic expressions for the decay constants, Eqs.~(\ref{eq:thedcs}).
We now take a derivative of each side with respect to $s'$ and take the limit $s'\rightarrow \infty$.
We may re-express the derivatives of the wave functions $\psi_{Vn}$
appearing in Eq.~(\ref{eq:intres1}) in terms of decay constants, for
$s=s'=\infty$. However, the sum is convergent only when one takes the limit $s' \rightarrow \infty$
with $s' \neq s$. The difference between the limit of the sum and its value at the limit point
is a divergent quantity $\Delta_V$, such that
\begin{equation}
\sum_n \frac{g_{Vn}^2}{q^2-m_{Vn}^2} + \Delta_V = \frac{3 U_{KK}^{-1/2} \kappa}{R^3} \left[
\frac{{U'}^{5/2} {\gamma'}^{-1/2}}{2s'} \partial_{s'} \psi_V^0(q^2,s') \right]^{s'\rightarrow \infty}  \,\,\, ,
\end{equation}
where primed quantities are functions of $s'$, or in original $U$ coordinates,
\begin{equation}
\sum_n \frac{g_{Vn}^2}{q^2-m_{Vn}^2}  + \Delta_V = \frac{3 U_{KK}^{-1/2} \kappa}{R^3} \left[
U^{5/2} \gamma^{-1/2} \partial_U \psi_V^0(q^2,U) \right]^{U \rightarrow \infty}  \,\,\, .
\label{eq:g2m2res}
\end{equation}
Sums that involve higher powers of $q^2-m_{Vn}^2$ in the denominator
are convergent, which suggests that $\Delta_V$ is a
constant independent of $q^2$.  This can be shown by taking $q^2$ derivatives in Eq.~(\ref{eq:intres1}),
so that the sum is convergent and there is no ambiguity in the evaluation of the subsequent limits,
and comparing to the same $q^2$ derivatives of Eq.~(\ref{eq:g2m2res}).  We fix $\Delta_V$ by noting that
$\psi_V^0(0,U)$ is a constant, so that the right-hand-side of Eq.~(\ref{eq:g2m2res}) vanishes in the
limit $q^2=0$; it follows that
\begin{equation}
\Delta_V = \sum_ m \frac{g_{Vn}^2}{m_{Vn}^2}  \,\,\,.
\end{equation}
The left-hand-side of Eq.~(\ref{eq:g2m2res}) may now be combined to give precisely $\Pi_V(-q^2)$
defined in Eq.~(\ref{eq:Pi}),
\begin{equation}
\Pi_V(-q^2) = \sum_n\frac{g_{Vn}^2 \, q^2}{m_{Vn}^2(-q^2+m_{Vn}^2)}= -\frac{3 U_{KK}^{-1/2} \kappa}{R^3} \left[
U^{5/2} \gamma^{-1/2} \partial_U \psi_V^0(q^2,U) \right]^{U \rightarrow \infty}  \,\,\, .
\label{eq:bigpiVres}
\end{equation}

To obtain the axial-vector self-energy, $\Pi_A(-q^2)$, one must take into account that
the axial bulk-to-boundary propagator $\psi_A^0(q^2,s)$ satisfies the boundary conditions
$\psi_A^0(q^2,-\infty)=-1$ and $\psi_A^0(q^2,\infty)=1$.  One can then derive an expression of
the same form as Eq.~(\ref{eq:g2m2res}) with $\Delta_V$, $g_{Vn}$, $m_{Vn}$ and $\psi_V^0(q^2,U)$ replaced
by $\Delta_A$, $g_{An}$, $m_{An}$ and $\psi_A^0(q^2,U)$, respectively.  The evaluation of $\Delta_A$,
however, is different in this case.  One can show that the right-hand-side of the axial
version of Eq.~(\ref{eq:g2m2res}) evaluates to a non-vanishing constant when
$q^2=0$, which
we identify with the square of the pion decay constant,
$f_\pi^2$.  Indeed, we find
that when $U_0=U_{KK}$ this constant agrees with the prediction
$f_\pi^2=4\kappa\,M_{KK}^2/\pi $ as determined in Refs.~\cite{SS1,SS2}
by studying the pion part of the effective action.  Thus we set
\begin{equation}
\Delta_A = \sum_ m \frac{g_{An}^2}{m_{An}^2} -f_\pi^2 \,\,\,,
\end{equation}
and
\begin{equation}
\Pi_A(-q^2) = -f_\pi^2+\sum_n\frac{g_{An}^2 \, q^2}{m_{An}^2(-q^2+m_{An}^2)} =
-\frac{3 U_{KK}^{-1/2} \kappa}{R^3} \left[
U^{5/2} \gamma^{-1/2} \partial_U \psi_A^0(q^2,U) \right]^{U \rightarrow \infty}  \,\,\, .
\label{eq:bigpiAres}
\end{equation}

\begin{figure}
\includegraphics[scale=.6]{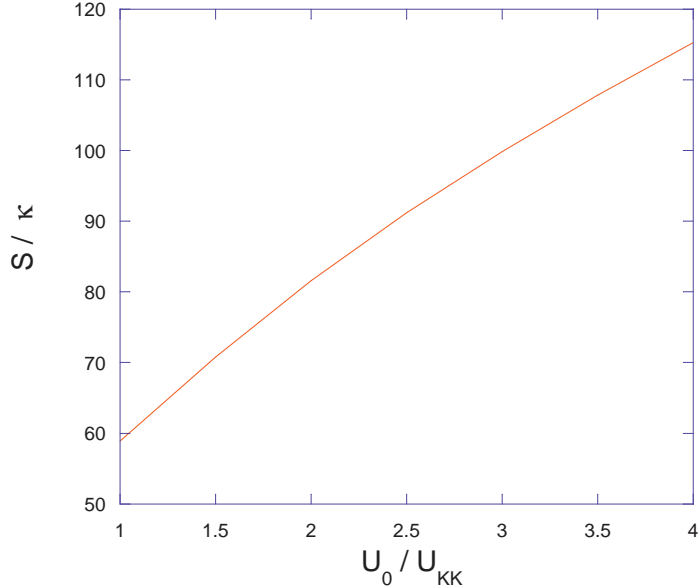}
\caption{\label{fig:S} Sum over all contributions to $S$ from the technicolor
sector, as a function of $U_0/U_{KK}$.}
\end{figure}

Using the definition of the $S$ parameter given in Eq.~(\ref{eq:S}), it immediately
follows that
\begin{equation}
S= - 12 \pi \frac{U_{KK}^{-1/2} \, \kappa}{R^3} \left[
U^{5/2} \gamma^{-1/2} \frac{d}{dq^2}\left(\partial_U \psi_V^0 - \partial_U \psi_A^0
\right) \right]^{U=\infty,\,\,q^2=0}  \,\,\,.
\label{eq:sfull}
\end{equation}
This result represents a sum over all resonances. We find that Eq.~(\ref{eq:sfull}) leads
to large, positive $S$ that increases with $U_0/U_{KK}$, a surprising result given the behavior
shown in Fig.~\ref{fig:S4}.
We explore this issue further in the next section. For the
antipodal D8-\Dbar configuration we obtain
\begin{equation}
S= 58.9 \, \kappa \approx 0.9 \,\,\,,
\end{equation}
where have again taken $g^2 N = 4 \pi$ and $N=4$ for the numerical evaluation.
This is a factor of $2$ larger than the estimate of $S$ in a one-doublet, SU(4)
technicolor models with QCD-like dynamics~\cite{PT}.  As a function of $U_0/U_{KK}$
we plot the $S$ parameter in Fig.~\ref{fig:S}.

It is worth noting that the scale $M_{KK}$, with respect to which we have been measuring masses and
decay constants, is determined by the electroweak scale
\begin{equation}
f_\pi^2=-\Pi_A(0)=(246\ {\rm GeV})^2,
\label{eq:fPiPiA}\end{equation}
where $\Pi_A(-q^2)$ is given by Eq.~(\ref{eq:bigpiAres}).
Eq.~(\ref{eq:fPiPiA}) is simply
one of the Weinberg sum rules with the renormalization condition
$\Pi_V(0)=0$, as is satisfied by Eq.~(\ref{eq:bigpiVres}).  This expression for
$f_\pi$ can also be derived by studying the two-point coupling of the axial
vector source ${\cal A}_\mu$ to the pion field, as in Ref.~\cite{SS2}.  As noted
earlier, we have set the pion field to zero in our analysis.
In the model with
$U_0=U_{KK}$ we find that $f_\pi^2=1.27 \,\kappa \,M_{KK}^2$, in agreement with Refs.~\cite{SS1,SS2}.
Hence, in that model $M_{KK}=218 / \sqrt{\kappa}$ GeV $\approx 1.8$ TeV, where in the last
approximation we set $g^2N=4\pi$ and $N=4$.  $M_{KK}$  sets the scale of additional Kaluza-Klein
states in the spectrum.  $M_{KK}$ also sets the scale of the technihadrons: in the same approximation
we find that the lightest technirho has mass $m_\rho\approx 1.5$ TeV; and the lightest techni-axial
vector has mass $m_{a_1}\approx 2.2$ TeV.  Note that once the electroweak scale is fixed, the only
remaining freedom in the model in addition to $N$ and $g^2N$
is the one-parameter choice of D8-brane configuration.  Although we
have found that the $S$ parameter is too large in this model to be a consistent model of EWSB, the
couplings between technicolor resonances and additional phenomenology can be calculated
in the model to the same accuracy as the calculations above. We turn our attention to ways in
which the $S$ parameter prediction of our model might be reduced in the next
section.

It is also worth noting that we can derive the above relations for $\Pi_V(-q^2)$
and $\Pi_A(-q^2)$ by the usual AdS/CFT prescription of
varying the action with respect to the sources for the currents ${\cal V}_\mu$ and
${\cal A}_\mu$.
In particular, the term in $\langle J_V^\mu(q)J_V^\nu(-q)\rangle$ proportional
to $g^{\mu\nu}$ is obtained by varying $iS_{{\rm source}}$, with
$S_{{\rm source}}$ given in
Eq.~(\ref{eq:Ssource}), with respect to $i{\cal V}_\mu$ twice, using
the equations of motion.  (The transverse
tensor structure can be obtained by keeping track of gauge fixing terms, as in
\cite{AdSQCD3}.)
The result is, \begin{equation}
\Pi_V(-q^2)=q^2 a_{V0}(q^2). \end{equation}
Comparing with Eq.~(\ref{eq:aV0aA0}) we see that the AdS/CFT procedure has successfully
reproduced our result, Eq.~(\ref{eq:bigpiVres}).  Similarly, we find
$\Pi_A(-q^2)=q^2 a_{A0}(q^2)$.

\section{Discussion}\label{sec:Conclusions}
\begin{figure}
\includegraphics[scale=.75]{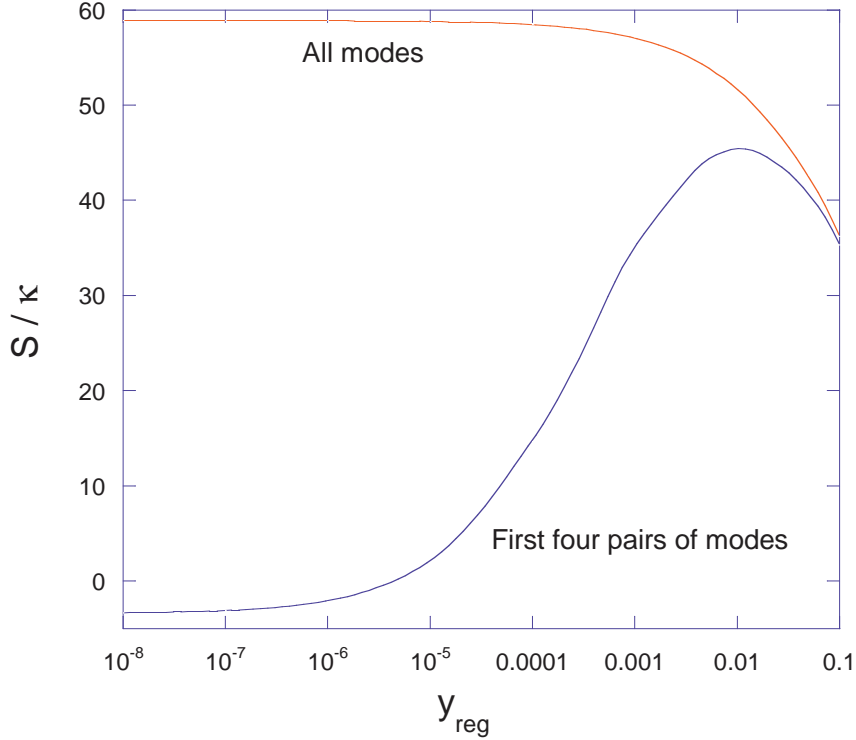}
\caption{\label{fig:UVreg} $S$ parameter as function of UV regulator $y_{reg}$.
As $y_{reg}$ increases, the $S$ parameter decreases and the higher resonances
decouple more quickly.}
\end{figure}

If we allow the D8-brane boundary to move in from infinity
artificially, then the confining scale can be raised as in
Refs.~\cite{CET,agashe}. In doing so, however, we are leaving the
realm of string theory in favor of a bottom-up approach to
holographic model building.  Models of electroweak symmetry
breaking constructed in this way have been discussed previously in
Refs.~\cite{Holo-EWSB1,Holo-EWSB2,CET,piai2}.  In our case, we may
consider a 5D model motivated by the AdS/CFT correspondence, but
with bulk geometry described by the induced metric on the
D8-branes in the D4-D8 system. The location of the UV boundary
determines the size of the radial dimension and, hence, the
resonance mass scale. Since the higher resonances decouple more
quickly as the size of the radial dimension is reduced, we can
test the agreement of the sum over modes with the sum rule
approach. Fig.~\ref{fig:UVreg} demonstrates the behavior of the
$S$ parameter as the boundary location is varied. Here we work in
the limit $U_0=U_{KK}$, and map the $U$ coordinate to a finite
interval $-1+y_{reg} < y < 1-y_{reg}$, via the relations
\begin{equation}
U = (U_{KK}^3 + U_{KK} z^2)^{1/3} \,\,\,\,\, \mbox{ and } \,\,\,\,\,
y = \frac{2}{\pi} \arctan(z/U_{KK}) \,\,\,.
\end{equation}
The physical boundaries at $U=\infty$ correspond to $y=\pm 1$. It is clear from
Fig.~\ref{fig:UVreg} that the lightest four pairs of vector and axial-vector modes
provide a better approximation of the $S$ parameter as $y_{reg}$ increases.
We also note that for a severely regulated theory, $y_{reg} \approx 0.9$, the $S$
parameter can be made consistent with the current limit $S \alt 0.09$~\cite{pdg}.
This corresponds to a radial dimension of proper length approximately $0.5/
M_{KK}$.   For the value $M_{KK}\approx 1.8$ TeV obtained in the previous
section, the radial dimension has size approximately (3.6 TeV)$^{-1}$.

It is interesting to note that this version of holographic duality for chiral
symmetries, with two boundary components rather than the single boundary
of Anti-de Sitter space, was anticipated by Son and Stephanov in Ref.~\cite{Son-S}.
They considered a model of mesons with light-quark quantum numbers based on an extension
of the notion of hidden local symmetry~\cite{Bando}.  The model is equivalent to a
deconstructed extra-dimensional SU$(N)$ gauge theory with an SU$(N_f)\times$SU$(N_f)$
flavor symmetry corresponding to the global symmetries of the fields localized at
the two boundaries.  To calculate current correlators, the global symmetries are weakly
gauged, and the gauge fields at the ends of the extra dimension act as sources for the
currents.  The configurations with SU$(N_f)$ gauge fields turned on at one boundary or
the other are analogous to the non-normalizable modes localized at the boundary of the D8 brane
or \Dbar brane segment.

A deconstructed~\cite{decons} version of the Sakai-Sugimoto model
would truncate the model to a finite number of resonances. Since
we have found that the light resonances in the Sakai-Sugimoto
model can contribute negatively to the $S$-parameter, it would be
interesting to study the behavior of the $S$-parameter as a
function of the number of lattice sites of the deconstructed
model.  Again, such an approach deviates from string theory since
there is no simple deconstruction of the radial dimension that
captures all of gravitational physics of the original theory, even
in the limit of a large number of lattice sites. Nevertheless,
deconstruction often leads to 4D theories that possess important
features of the higher-dimensional theory that they are designed
to approximate.  In the present case, the Sakai-Sugimoto model
would simply provide a paradigm for the choice of gauge groups and
symmetry breaking pattern in the deconstructed theory. Electroweak
constraints would require detailed study, especially away from the
large $N$ limit where radiative corrections are generally
non-negligible and must be taken into account.  It would also be
interesting to compare these deconstructed models with existing
models of electroweak symmetry breaking that were motivated in
part by deconstruction of extra dimensions, such as little Higgs
models~\cite{littleHiggs}.

In summary, we have studied the viability of the Sakai-Sugimoto model of intersecting D4 and
D8 branes as a technicolor-like model of electroweak symmetry breaking.
We determined the masses of the technicolor resonances and their decay
constants in terms of the electroweak scale.
We found that the contribution of the strong dynamics to the Peskin-Takeuchi $S$ parameter is
larger by at least an order of magnitude than the experimentally allowed value $S\alt 0.09$, even allowing for
adjustment of the D8-brane configuration.  However, we observed that the lightest pair of vector and
axial vector resonances contributes negatively to $S$, while the final result reflects a
non-decoupling of heavy resonances that seems to be sensitive to the details of the model. The model
can be made consistent with precision electroweak constraints by artificially introducing an ultraviolet
regulator, as we showed in Fig.~\ref{fig:UVreg}, or perhaps by deconstructing the model along the radial
direction in order to truncate it to the lightest several technicolor resonances.  In order to become a
complete model of electroweak symmetry breaking, the Standard Model fermions
would need to be included together with a mechanism for generation of their
masses.

%%%%%%%%%%%%%%%%%%%%%%%%%%%%%%%%%%%%%%%%%%%%%%%%%%%%%%%%%%%%%%%%%%%%
\begin{acknowledgments}
We thank Ofer Aharony and Veronica Sanz for useful discussions.  CDC thanks the NSF for support under
Grant No.~PHY-0456525.  M.S. thanks the NSF for support under Grant No.~PHY-0554854. The work
of J.E. is supported in part by the NSF under Grant No.~PHY-0504442 and the
Jeffress Foundation under Grant No.~J-768.
\end{acknowledgments}
%%%%%%%%%%%%%%%%%%%%%%%%%%%%%%%%%%%%%%%%%%%%%%%%%%%%%%%%%%%%%%%%%%%%

\end{document}